\documentclass{sig-alternate-05-2015}

\usepackage{graphicx}
\usepackage[space]{grffile}
\usepackage{latexsym}
\usepackage{textcomp}
\usepackage{longtable}
\usepackage{multirow,booktabs}
\usepackage{amsfonts,amsmath,amssymb}
\usepackage[numbers]{natbib}
\usepackage{url}
\usepackage{hyperref}
\hypersetup{colorlinks=false,pdfborder={0 0 0}}
\newif\iflatexml\latexmlfalse
\usepackage[utf8]{inputenc}
\usepackage[english]{babel}
\newcommand{\hz}{\ensuremath{~\mathrm{Hz}}}
\newcommand{\CS}{\ensuremath{\boldsymbol{\mathrm{CS}_1}}}
\newcommand{\Run}{\ensuremath{\boldsymbol{\mathrm{Run}_{1440}}}}
\newcommand{\PCo}{\ensuremath{\mathrm{PC}_1}}
\newcommand{\PCt}{\ensuremath{\mathrm{PC}_2}}

\def\sharedaffiliation{%
\end{tabular}
\begin{tabular}{c}}

\begin{document}


\conferenceinfo{BuildSys 2016}{November 16-17, 2016, Stanford, CA, USA}

\title{Short paper: Hypertemporal Imaging of NYC Grid Dynamics}

%
%
%
%
%

\numberofauthors{4}
\author{
%
%
\alignauthor
Federica B. Bianco\\
\affaddr{CUSP \& Center for Cosmology and Particle Physics, NYU}  \\
\email{fbianco@nyu.edu}
\alignauthor
Steven E. Koonin\\
\affaddr{CUSP}\\
\email{sek9@nyu.edu}
\alignauthor
Charlie Mydlarz\\
\affaddr{CUSP \& Music and Audio Research Laboratory, NYU}  \\
\email{cm3580@nyu.edu}
\alignauthor
\and
Mohit S. Sharma\\
\affaddr{CUSP}\\
\email{mohit.sharma@nyu.edu}
\sharedaffiliation
\affaddr{CUSP: Center for Urban Science and Progress}  \\
\affaddr{New York University}  \\      
\affaddr{1 MetroTech Center, 19th Floor}  \\
\affaddr{Brooklyn, NY 11201}
}

\CopyrightYear{2016} 
\setcopyright{acmlicensed}
\conferenceinfo{BuildSys '16,}{November 16 - 17, 2016, Palo Alto, CA, USA}
\isbn{978-1-4503-4264-3/16/11}\acmPrice{\$15.00}
\doi{http://dx.doi.org/10.1145/2993422.2993570}

\maketitle


\begin{abstract}

Hypertemporal visible imaging of an urban lightscape can reveal the phase of the electrical grid granular to individual housing units. In contrast to \emph{in-situ} monitoring or metering, this method offers broad, persistent, real-time, and non-permissive coverage through a single camera sited at an urban vantage point.  Rapid changes in the phase of individual housing units signal changes in load (\emph{e.g.}, appliances turning on and off), while slower building- or neighborhood-level changes can indicate the health of distribution transformers. We demonstrate the concept by observing the 120\hz~flicker of lights across a NYC skyline.  A liquid crystal shutter driven at 119.75\hz~down-converts the flicker to 0.25\hz, which is imaged at a 4\hz~cadence by an inexpensive CCD camera;  the grid phase of each source is determined by analysis of its sinusoidal light curve over an imaging ``burst'' of some 25 seconds.  Analysis of bursts taken at $\sim15$ minute  cadence over several hours demonstrates both the stability and variation of phases of halogen, incandescent, and some fluorescent lights.  Correlation of such results with ground-truth data will validate a method that could be applied to better monitor electricity consumption and distribution in both developed and developing cities.

\end{abstract}

\ccsdesc[500]{Hardware~Energy metering}
\ccsdesc[500]{Hardware~Emerging tools and methodologies}
\ccsdesc[300]{Hardware~Circuits power issues}
\ccsdesc[300]{Hardware~Aging of circuits and systems}
\ccsdesc[300]{Hardware~Transient errors and upsets}
\ccsdesc[300]{Hardware~Power networks}

\printccsdesc
\keywords{Urban science: observing techniques; imaging; time series analysis.}

%
\bibliographystyle{abbrv}
%




\section{Introduction}
The 60\hz~AC line frequency of an electrical grid in the US induces a 120\hz~flicker in most of lights that it powers, including incandescent, halogen, transitional fluorescent, and some LED sources.\footnote{Modern fluorescent lights are ballasted electronically, rather than magnetically, and flicker at 5--40~kHz.} This flicker is generally imperceptible to the unaided eye.

The 60\hz~line frequency is universal across the grid and maintained to within $\sim0.02\hz$,\footnote{\href{http://fnetpublic.utk.edu/index.html}{http://fnetpublic.utk.edu/index.html}} but the phase of the voltage driving any particular light (and hence of its flicker) will depend upon the grid’s generating sources, topology and condition of its reactive components (\emph{e.g.}, distribution transformers), and the load local to the light.  Determination of such phases and their time variation can therefore probe grid dynamics on multiple temporal and spatial scales.
A single camera at an urban vantage point (\emph{e.g.}, the roof of a tall building) can persistently observe 10’s of thousands of lights in thousands of buildings~\citep{Dobler2015}. The individual phases of many lights could then be determined straightforwardly, simultaneously, and non-permissively by high-speed imaging photometry.   However, the requirements of high cadence ($\sim10^3\hz$), wide angle, high sensitivity, and high spatial resolution imply expensive equipment and large bandwidth, data storage, and computational capabilities. High-speed, low-light condition cameras (\emph{e.g.} EMCCDs: Electron Multiplying Charge Coupled Devices) 
are typically expensive and have small fields of view. Instead, we have chosen to chop the image at near-line frequency (119.75\hz) with a liquid-crystal shutter and thus down-convert the flicker to a beat frequency of $\sim0.25\hz$, which is then easily imaged at $4\hz$ cadence with a small digital camera.

This paper presents a proof-of-concept for observing grid phase by persistent and synoptic visible imaging of flickering city lights. We analyze a dataset of 25-second bursts of 4\hz~down-converted images of a New York City skyline acquired over the course of two hours. Analysis of these data shows the expected 0.25\hz~beat frequency and reveals the presence of both stable and varying phase shifts among some 50-100 light sources; we attribute the latter to changes in the grid during the 1-hour observing period. In the spirit of reproducibility and open science all the code used to generate the results and plots presented here is collected in the GitHub repository \href{https://github.com/fedhere/detect120/}{https://github.com/fedhere/detect120/}, where interactive versions of the figures in this paper and additional figures can also be found.

\section{Equipment}\label{sec:eq}
Our system includes a camera and a shutter. The camera is a PointGrey Flea3 5.0~MP Color GigE Vision instrument 
capable of frame rates of up to 8\hz~on the full field of view, which is imaged on a $2448\times2048$ pixel CCD (SONY ICX655) in 3 color channels. The camera is controlled by the manufacturer’s proprietary software from a laptop, and is equipped with a 35mm f/2.0-16 2/3'' 10~MP lens and coupled with a liquid crystal shutter mounted at the lens aperture.
The shutter (ThorLab LCC1620
) attenuates visible wavelengths (420–700~nm) through an optically active liquid crystal cell flanked by polarizers. We operate the shutter in \emph{shutter mode} driving it with a 5V pure audio tone that induces the transmission to vary from a maximum opacity of 60\%-80\% (depending on the wavelength) and a minimum opacity $\sim10\%$.
Nonlinearities in the shutter response cause an opening time $\sim5$~ms and a closing time $\sim1$~ms, times that are a significant fraction of the $\sim8.3$~ms flicker. 
We have studied the system response by observing a single incandescent light bulb in the laboratory (to be reported in a separate publication) and have demonstrated that the slow shutter response does not materially affect our ability to detect or analyze the down-converted flicker.

\section{Data}\label{sec:data}
Our goal is to detect both secular changes in phase that may be due to transformer degradation and episodic shifts in phase that may occur over a few seconds   due to changes in load or generation. We expect the latter to be both rare and small, so that real-time detection would require accurate continuous measurements over hours.  Instead, we monitor the same city scene at regular, short intervals, during the course of a night. Every 5-to-15 minutes we image 2-to-5 minutes of the scene at a 4\hz~cadence, with an exposure time ($\lesssim125$~ms) that varies to maximize the signal-to-noise ratio (SNR). 
Here we report the study of a single dataset.

\underline{ City scene 1, May 2016 (\CS):}
data taken with a shutter speed of 119.75\hz, comprising five 5-minute runs of 4\hz, 100~ms exposures (1,200 images) separated by $\sim15$ minutes. Each run is  flanked by 20 longer exposures of 250~ms (limited by the camera software) with the shutter fully transparent (which we call \emph{shutter-free} sequences). These are stacked to obtain deep images from which the light sources are identified (\autoref{sec:lightselection}). We have analyzed a burst of 100 consecutive images (25 seconds) from each run to determine grid phases for each source and their inter-burst variation.

{\underline{ \Run:}}
A subset of the \CS~dataset: six 25-second bursts (100 consecutive images), with an intra-burst cadence of 37.5 seconds. These data were used to create and test the data reduction pipeline and to assess the short term stability of both our system and the extracted phases.

\begin{figure}[h!]
\begin{center}
  \includegraphics[width=0.7\columnwidth]{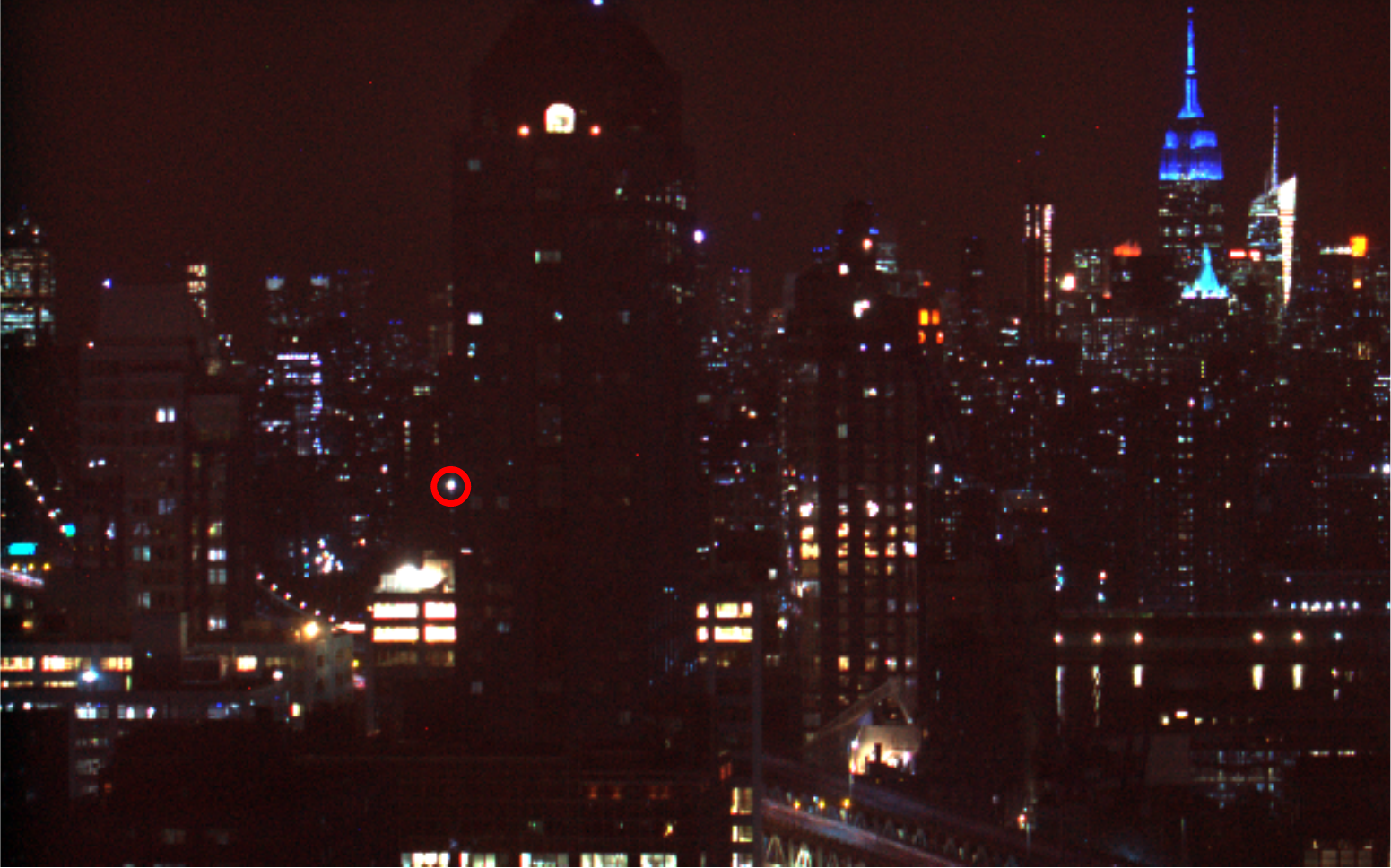}
  \end{center}
\caption{\label{fig:stacks} RGB stacked image generated by taking the median, pixel by pixel, of 20 200~ms \emph{shutter-free} images.  The view is of Northern Brooklyn and the Lower East Side of Manhattan. The Empire State Building, seen in blue in the upper right, is approximately 3.8 miles away. We isolate 1,533 sources in this scene. The source circled in red was chosen as the phase reference (Section~\ref{sec:results}). Its lightcurve during a 25 seconds burst is shown in Figure~\ref{fig:PCA}.}
\end{figure}

The scene imaged in our data 
is shown in Figure \ref{fig:stacks}.

\section{Analysis}\label{sec:analysis}

\subsection{Source Identification}\label{sec:lightselection}
Our automated algorithm for identifying and locating individual sources in the images begins by stacking the shutter-free images to build a median image.  Contiguous regions of pixels that exceed a well-chosen threshold are then identified as sources (either windows or exterior lights).
If needed, images in a stack and bursts of images are registered in Fourier space (using the package \href{https://github.com/fedhere/coaddfitim}{\tt{coaddfitim}}). However, with the mechanical stability of our camera mount images over a 2-hour observing session typically do not require registration.

Light sources are automatically selected from the median stacked image by thresholding a high pass filtered version of the stack in a process that includes the following steps:\footnote{Associated code (\href{https://github.com/fedhere/detect120/blob/master/stackimages.py}{\tt{stackImages.py}} and \href{http://github.com/fedhere/detect120/blob/master/windowFinder.py}{\tt{windowFinder.py}}) can be found in the project's \href{https://github.com/fedhere/detect120}{GitHub repository}.}

\vspace{-1em}

\begin{enumerate}
  \itemsep-0.45em 
\item{The stack is smoothed with a circular Gaussian kernel with a standard deviation of 10 pixels. This smoothed image is then subtracted, pixel by pixel, from the original stack to produce a high-pass image.}
  
\item{The $90^\mathrm{th}$ percentile of the distribution of pixel values in the high-pass image is taken to be the selection threshold: pixel with values above the threshold are retained as illuminated pixels.}
  
\item{The illuminated pixels selected in Step 2 are merged into a ``patch'' when adjacent, and patches larger than $10\times10$ pixels are identified as ``sources''.}
\end{enumerate}

\vspace{-1em}

In the data presented here, this process identifies 1,533 sources
which we monitor to detect 0.25\hz~oscillations. During bursts, the lightcurve (time series of light intensity) of each source is extracted by summing the pixels in a $2\times2$ square aperture at the center of brightness of its patch.\footnote{Our aperture is far smaller than the 100 pixel limit observed by the CUSP Urban Observatory to preserve privacy~\citep{Dobler2015}.}

\subsection{Source Selection}\label{sec:PCA}
\begin{figure}[h!]
\begin{center}
  \includegraphics[width=0.8\columnwidth]{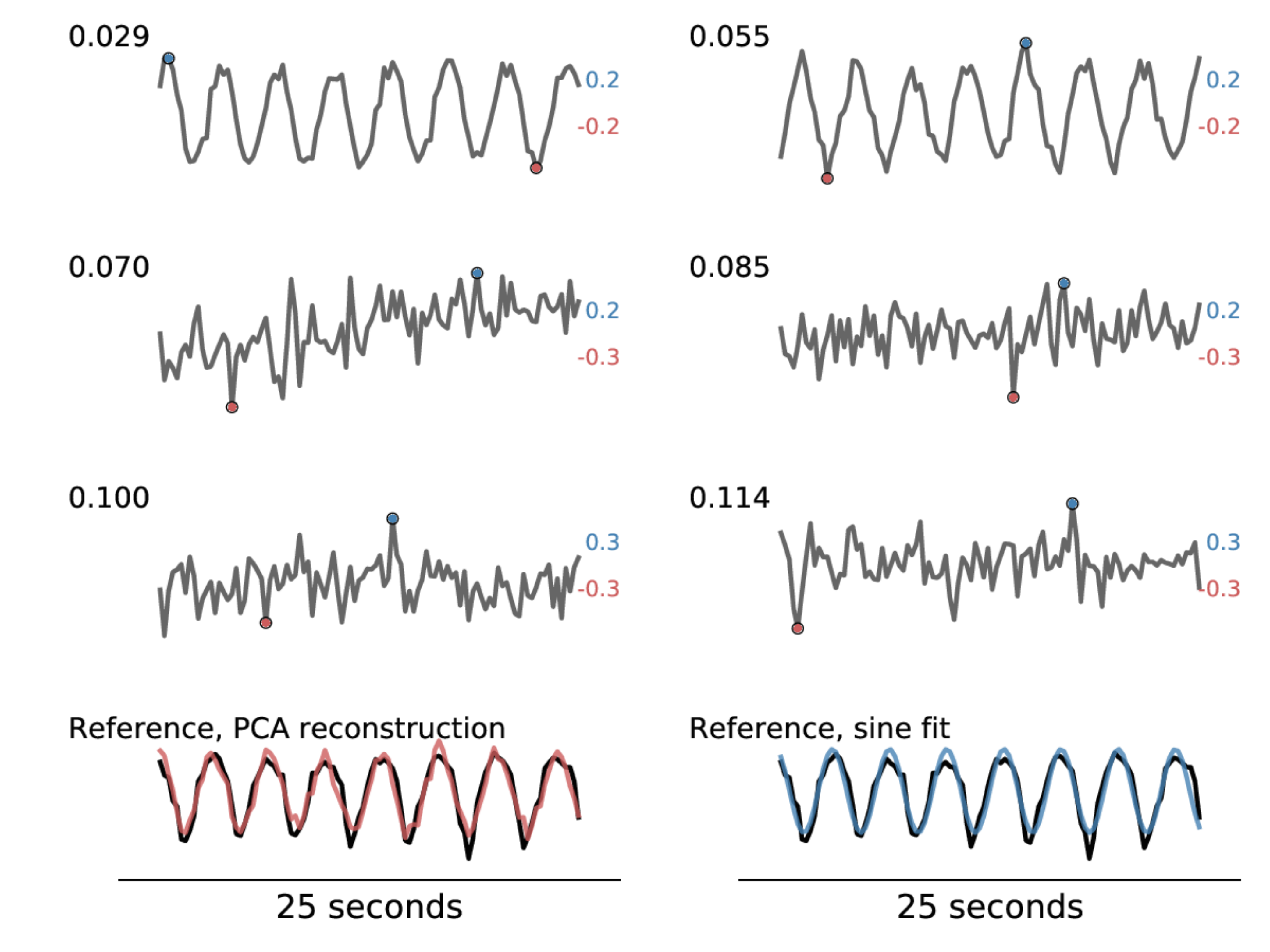}
\caption{\label{fig:PCA} PCA decomposition of 4\hz~lightcurves over a 25-second burst of \Run.  The first six orthonormal principal components are displayed left to right and top to bottom in order of their ability to explain the variance of the data. At the top left of each panel the cumulative fraction of the variance explained up to that component is shown. To the right of each panel the minimum (red) and maximum (blue) of that component are reported. The first two components, \PCo~and \PCt, describe the expected $\sim0.25\hz$ frequency of the down-converted 120\hz~flicker, but together explain only $\sim6\%$ of the variance. The last two panels show the lightcurve of the reference source (Figure~\ref{fig:stacks}) in black (left and right panel). We overplot its reconstruction with \PCo~and \PCt~(left, red), and the sine fit to the lightcurve (right, blue), from which we determine $\phi$ (with a 68\% confidence intervals of [+0.064,-0.065] radians),  and $\nu = 0.29^{+0.002}_{-0.002}$\hz.}
\end{center}
\end{figure}

The 4\hz~imaging of the sources yields diverse lightcurves, but a principal component analysis (PCA\footnote{PCA is performed using the {\tt scikit-learn} Python package {\tt PCA} module, lightcurve selection and extraction with \href{https://github.com/fedhere/detect120/blob/master/getalllcvPCA.py}{\tt{getalllcvPCA.py}}, available in the project's \href{https://github.com/fedhere/detect120}{GitHub repository.}}, \cite{Pearson_1901}, \cite{Jolliffe_1986}), which projects them onto an optimal orthonormal basis, consistently identifies the two most important principal components (those that explain the largest and second largest fractions of the lightcurve set variance) to be nearly-sinusoidal curves (\emph{pseudosine} hereafter) with the expected beat frequency $\sim0.25\hz$.

The first six principal components of a 25-second burst from \Run~are shown in \autoref{fig:PCA}. The first two (\PCo~and \PCt) are pseudosines in quadrature, as is required to reconstruct a pseudosine of arbitrary phase.  The reconstruction of the lightcurve of our reference source through \PCo~and \PCt~is also shown (bottom, left). However, these first two components typically account for only $\sim0.05 - 0.3$ of the variation in the lightcurves; 82 components are needed to explain  90\% of the variance for this particular burst. While flicker is clearly present in, and sometimes dominates, the signal for a fraction of the sources, the remainder show diverse behavior. We attribute this ``noise'' to atmospheric turbulence, timing imperfections in the shutter and camera, and lighting technologies that do not flicker at 120\hz.

\begin{figure}[h!]
\begin{center}
  \includegraphics[width=0.75\columnwidth]{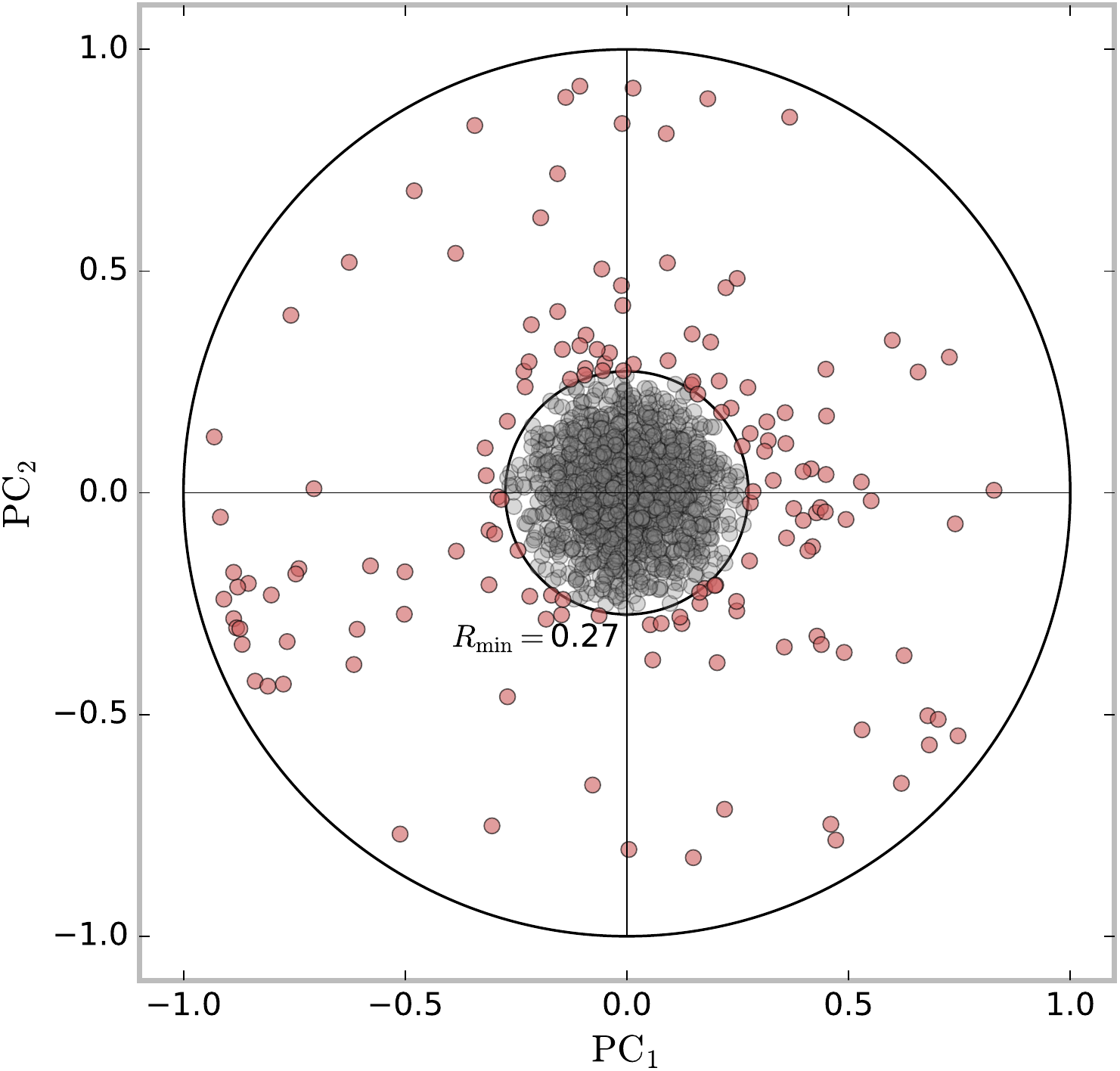}
  \caption{\label{fig:PCAselection} PCA-based selection of lightcurves: the lightcurves' projection on the \PCo-\PCt~plane is shown for the 25-seconds burst described in Figure~\ref{fig:PCA}. A lightcurve's distance from the origin, $R$, is a measure of the importance of the pseudosine components in describing the source's behavior. The inner circle is the selection threshold, $R_\mathrm{min}=0.27$, which isolates that 10\% of lightcurves (red dots) showing the pseudosine structure most prominently.}
\end{center}
\end{figure}

The fraction of variance spanned by the first two components is affected by the window selection -- a more liberal window selection that allows low SNR sources in the dataset shows a larger variation not captured by the pseudosine signal. The same two principal components, however, do capture a large portion of the dataset signal within the expected bandwidth of the pseudosines. Upon filtering the lightcurves with a Gaussian filter of width $\sim0.25\hz$ applied in Fourier space and centered on 0.25\hz~the first two components, still pseudosines in quadrature, explain $\sim25\%$ of the variance, and 18 components capture $>90\%$. 

The lightcurve of each source can be projected onto the two pseudosine components, and plotted in the plane defined by these two inner products (\PCo-\PCt~plane).  We use the distance from the origin in that plane, $R~=~\sqrt{\PCo^2 + \PCt^2} <1$, as the selection criterion to identify those lightcurves with the largest pseudosine contribution, and so select the top 10\% sources for further analysis (Figure~\ref{fig:PCAselection}).


\subsection{Phase determination}\label{sec:fitting}
Each of the selected lightcurves (153 for \Run) is fit with a sine of frequency $\nu$ and phase $\phi$. 
We 
compute the best-fit
values and confidence intervals of $\nu$ and $\phi$ by sampling the parameter
space with the \emph{emcee} Python implementation~\citep{ForemanMackey13} of the Affine Invariant Markov Chain Monte Carlo~\citep{GoodmanWeare}, defining the likelihood for the lightcurve $\pmb{l}(\pmb{t})$ as:
  $$
L(\nu,~\phi|~\pmb{l}) \propto \exp[-\chi^2(\nu,~\phi,~\pmb{l})/2] = \exp[-(\pmb{l} - \mathrm{sin}(\nu,~\phi,~\pmb{t}))^2/2]
  $$
 (ignoring the intrinsic observational uncertainties). 
  Over \Run, the uncertainty $\epsilon$ in the phase determination averages to $<\epsilon_\phi>=0.14\pm 0.016$, while the mean of the uncertainty in the frequency is $<\epsilon_\nu>=0.004\pm 0.0005\hz$.

For all \CS~data we measure beat frequencies $\nu\sim0.29\hz$, corresponding to a possible 0.035\% deviation from 120\hz~in the original frequency. This deviation is within the precision of our system and of the line frequency.

\section{Results}\label{sec:results}
In this work we present the analysis of six consecutive bursts from \Run~and five bursts from consecutive runs in \CS. Consistently across the dataset we identify a minimum of 50 sources with SNR sufficient to unambiguously determine the phase. For example: in six 100 images bursts in \Run, a quality cut on the parameters fit ($\chi^2(\nu,~\phi,~\pmb{l})<=1$) selects a minimum of 57, and a maximum of 70, sources).

Since the phases we determine are defined relative to the start of a burst and we do not have sufficient timing accuracy to maintain continuity between runs (or even bursts), we refer each phase to that determined for the reference source (Figures \ref{fig:stacks} and \ref{fig:PCA}), which has a consistently large $R$.

\begin{figure}[h!]
\begin{center}
  \centering
  \includegraphics[width=0.36\textwidth]{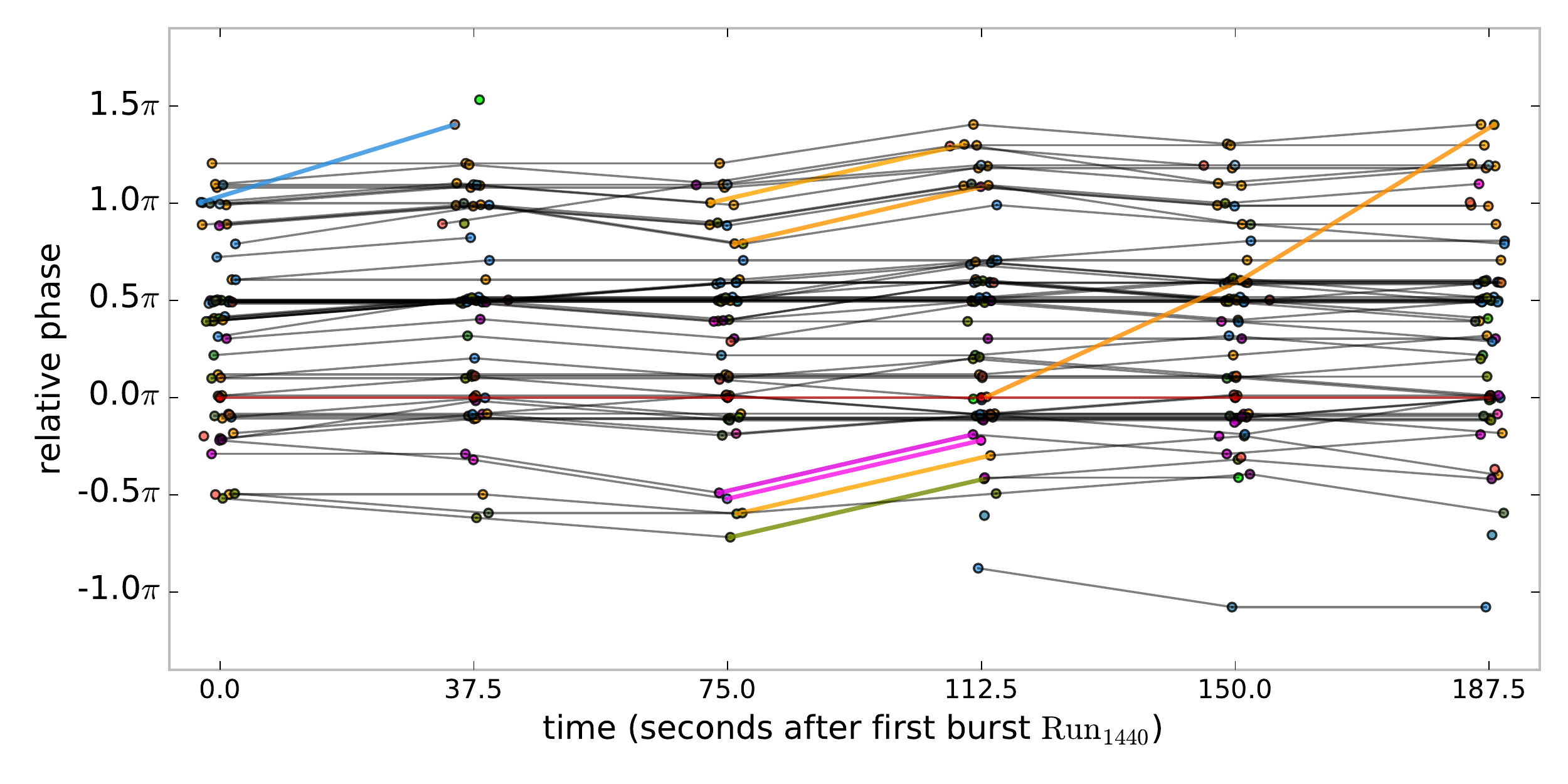}
  \includegraphics[width=0.36\textwidth]{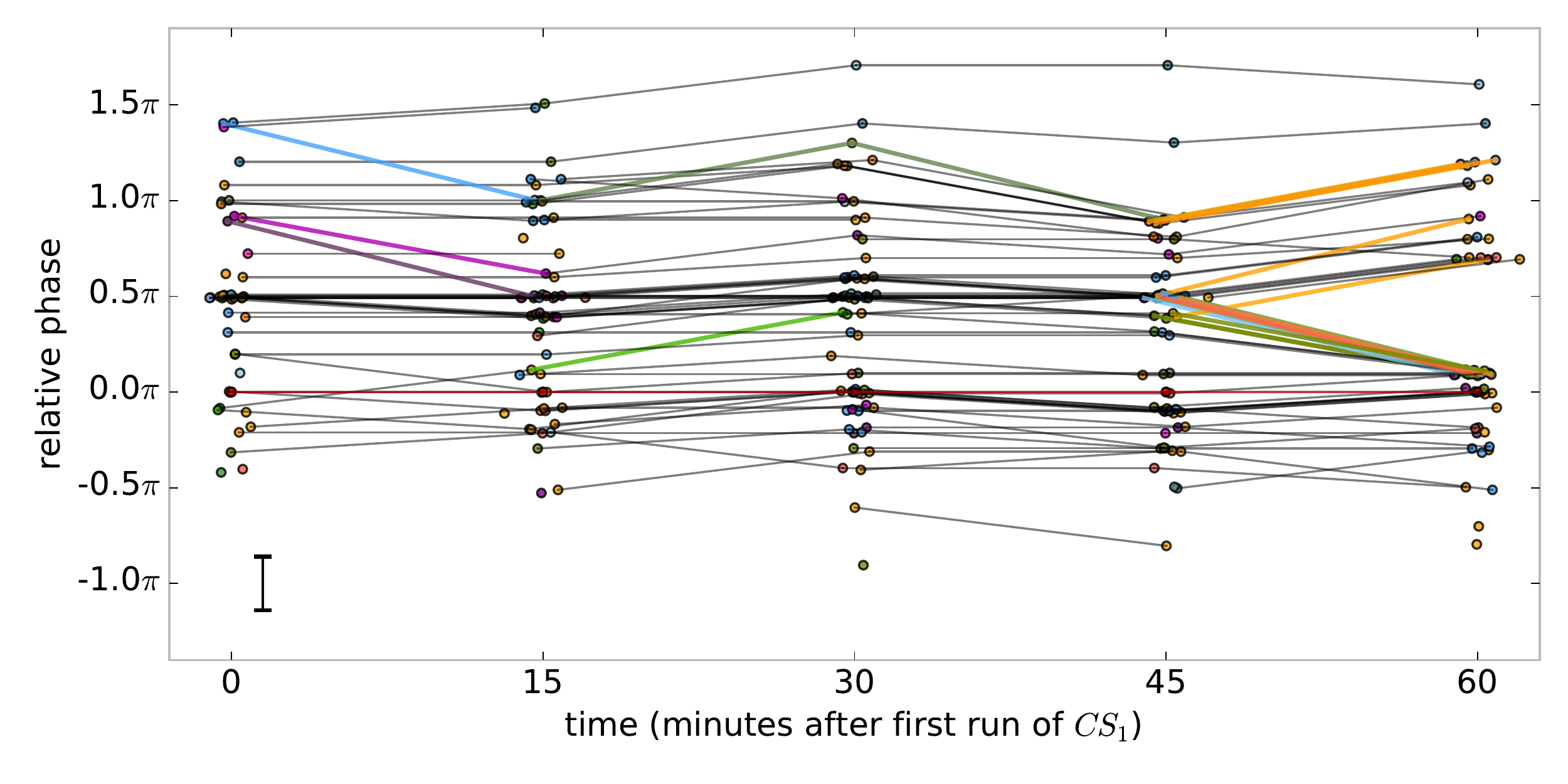}


  \caption{\label{fig:phases} Evolution of source phases relative to the reference within a burst (top) and between bursts
(bottom); phases are rounded to one 
decimal digit. The typical error bar, obtained from
sampling the parameter space (\autoref{sec:fitting}), is
indicated at the bottom-left of the lower plot.  The relative phase of each source at each time stamp is indicated by a colored dot; sources belonging to the same building
are plotted in the same color, with small offsets to
enhance readability. Changes in relative phase larger than $0.3\pi$
are shown with a similar color coding.}

\end{center}
\end{figure}

The evolution of the relative phases through the six bursts of \Run~is shown in the top panel of Figure~\ref{fig:phases} and changes over the five runs of \CS~are shown in the bottom panel; changes larger than $0.3\pi$ radians are highlighted by colored lines. We find general stability in the relative phases on  both minute and hour time scales, but also detect isolated changes larger than our observational uncertainties that are consistent with changing circuit loads.  

\begin{figure}[h!]
\begin{center}
  \centering
  \includegraphics[width=0.38\textwidth]{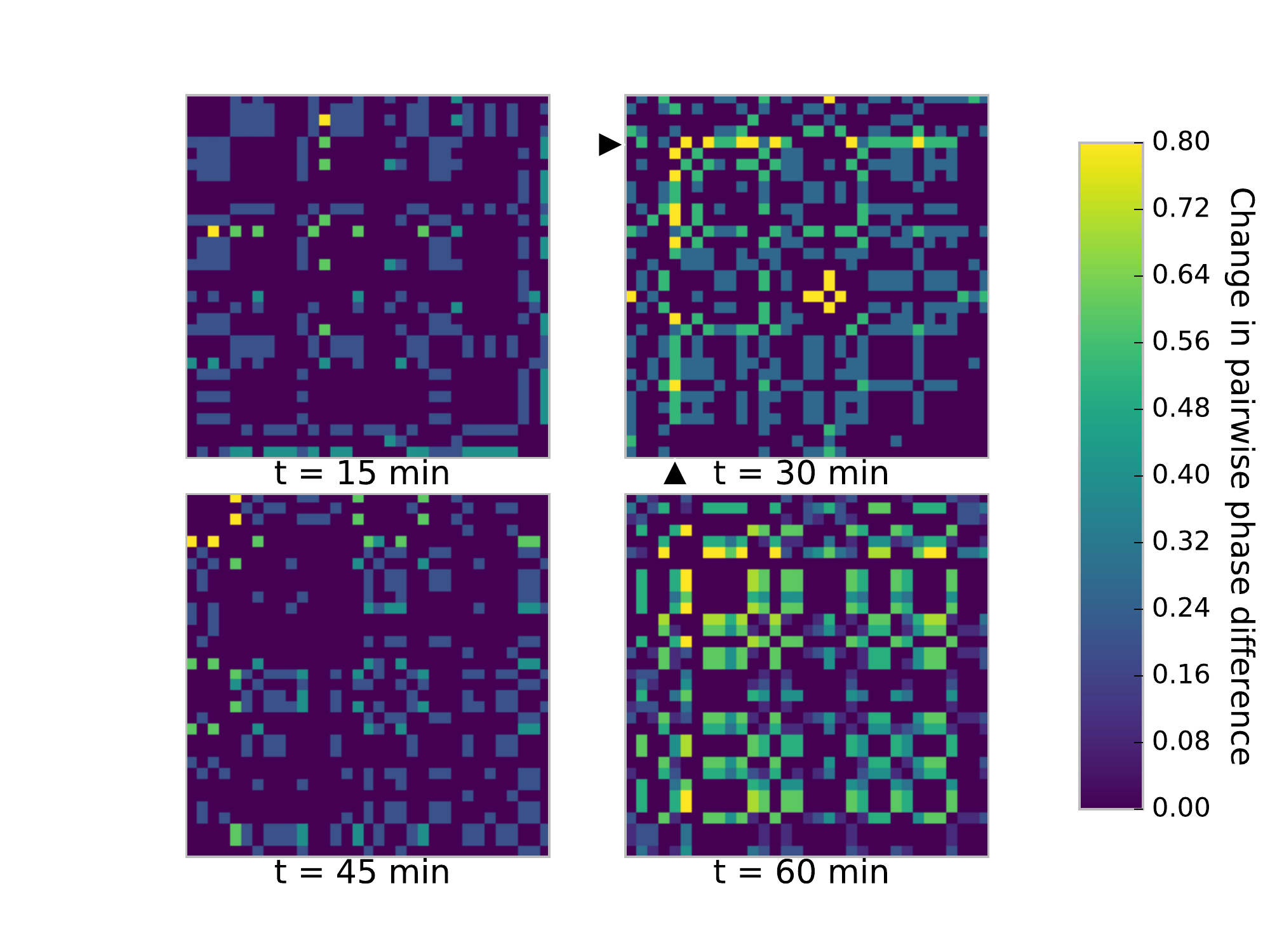}
  \caption{\label{fig:matrix} Phase changes between the runs shown in the lower panel of Figure 4.  Each panel represents the change in pairwise phase differences between bursts in successive runs.  Each row and column of a panel represents a source and the color of each pixel in the matrix indicates the change in pairwise phase difference from one run to the next. The rows and columns are ordered so that sources  in the same building are adjacent. An arrow highlights a changing phase between t=15 and t=30 minutes.}
     
\end{center}
\end{figure}

In Figure~\ref{fig:matrix} we show inter-run changes in the pairwise phase differences for all sources selected at high SNR consistently throughout \CS. For each pair of sources the color indicates the change in the pairwise phase differences relative to the previous time stamp. Subtle changes in the structure of these similarity matrices (\emph{e.g.} the fifth source between the 15 and 30 minutes time stamps, marked by an arrow in the second panel) indicate changes in the phase of a source. Sources are ordered so that those belonging to the same building are adjacent; changes affecting more than one row or column thus indicate sources on the same circuit.

In summary, we have demonstrated that hypertemporal imaging of an urban lightscape can broadly, persistently, and non-permissively detect the dynamics of the electrical grid on timescales ranging from minutes to hours with a granularity to the individual housing units. Our results motivate ongoing work to refine and exploit the methodology, including a larger camera/lens system to increase the SNR and allow more distant sources to be studied, more reliable timing of the camera and shutter system, field experiments to validate the ground truth, and systematic observations of phase changes which, when fused with correlative data on the electrical grid and the buildings observed, should reveal details about consumption patterns and grid status.

\noindent{\bf Acknowledgements:} This work made use of Python modules including \verb=Matplotlib=, \verb=NumPy=, \verb=scikit-learn=, \verb=SciPy=, and \verb=PythonImageLibrary=.
Some plots were
created with \verb=sparklpy=, 
DOI:10.5281/zenodo.35387, 
\url{https://github.com/fedhere/sparklpy}.

\bibliography{bibliography/converted_to_latex.bib%
}

\end{document}